\documentclass{mem}
\usepackage{natbib}\usepackage{txfonts}\usepackage{balance}
\usepackage{graphicx}
\usepackage[a4paper]{hyperref}
\idline{75}{282}

\begin{document}

\title{The ``double final fate'' of super-AGB stars and its possible consequences for some astrophysical issues}

\author{M.L. \,Pumo\inst{1,2} 
\and P. \,Ventura\inst{3}
\and F. D'Antona\inst{3}
\and R.A. \,Zappal\`a\inst{4}
}

\offprints{M.L. Pumo}

\institute{
CSFNSM, c/o Dip. di Fisica e Astronomia 
dell'Universit\`a di Catania, Via S. Sofia 64, 
I-95123 Catania, Italy,
\email{mlpumo@ct.astro.it}
\and
INAF - Osservatorio Astrofisico di Catania,
Via S. Sofia 78, I-95123 Catania, Italy
\and
INAF - Osservatorio Astronomico di Roma, 
Via Frascati 33, I-00127 Roma, Italy
\and
Universit\`a di Catania,
Dip. di Fisica e Astronomia, Via S. Sofia 78,
I-95123 Catania, Italy
}

\authorrunning{Pumo et al.}

\titlerunning{The ``double final fate'' of SAGBs stars and its possible consequences}
   \subtitle{}

\abstract{Super-AGB stars can conclude their evolution either as neon-oxygen white dwarfs or as electron-capture supernovae. We discuss the possible consequences of the existence of this ``double final fate'' in the self-enrichment of globular clusters and in the nucleosynthesis process of s-nuclei.

\keywords{Stars: AGB and post-AGB - globular clusters: general - Nuclear reactions, nucleosynthesis, abundances} }

\maketitle


\section{Introduction}

It is well known that stars can be divided into different groups characterised by the same final fate \citep[e.g.][]{Poel08}. Roughly speaking we consider two categories of stars: the first one includes the low- and intermediate-mass stars which end their evolution as white dwarfs, being not able to ignite carbon since they are less massive than $M_{up}$ ($\sim$ 7-9$M_{\odot}$) defined as the minimum initial mass for the carbon ignition \citep[e.g.][]{BI80}; the sencond one is formed by the massive stars that conclude their evolution as iron core collapse supernovae, being more massive than $M_{mas}$ ($\sim$ 11-13$M_{\odot}$) defined as the minimum initial mass for the completion of all the nuclear burning phases \citep[e.g.][]{woosley02}. 

In this scenario the class of stars with initial mass between $M_{up}$ and $M_{mas}$ is missing. These stars, referred to as super-AGB (SAGB) stars, can have a ``double final fate'' \citep[e.g.][]{sp06}. In fact, SAGB stars are massive enough to ignite carbon but, being unable to evolve through all nuclear burning stages, they conclude their evolution either forming a neon-oxygen white dwarf (NeO WD) or going through an electron-capture supernova (ecSN) becoming a neutron star, if electron-capture reactions are efficiently activated \citep[e.g.][]{ritosa99}. 

Recent studies \citep[][]{sp06,thesis,vienna06,s07,Poel08} have shown that both final evolutionary channels exist for metallicities ranging from Z=$10^{-5}$ to Z=$0.04$.

The existence of this double fate for SAGB stars could have important consequences in relation to different astrophysical problems such as issues associated with the mass distribution of white dwarfs, the properties of novae, the pulsar and supernova rates, as well as with the self-enrichment of globular clusters and the nucleosynthesis processes of trans-iron elements \citep*[see for details][PDV08 hereafter]{firenze07,s07,roma08}. We discuss their possible role in the last two above mentioned issues.


\section{The role of SAGB stars in the self-enrichment of globular clusters}

Recent works seem to indicate that at least four are the globular clusters (GCs) --- $\omega$~Cen, NGC~2808, NGC~6441 and NGC~6388 --- having a non negligible fraction (10-15\% of the total stellar population) of very helium rich (Y$\gtrsim$0.35) stars \citep[][]{norris2004,dantona2005,piotto2005,piotto2007,caloi-dantona2007a}. Such stars can be intepreted in terms of a ``second-generation'' stellar population originated through the so-called self-enrichment process from the helium-rich ejecta of stars belonging to a previous stellar generation \citep[e.g.][]{VD05}, but the nature of the progenitors having the required high helium abundance in their ejecta is still an open question (e.g. PDV08 and references therein). Either massive AGB stars subject to hot bottom burning \citep[][]{ventura2001, ventura2002} or fastly rotating massive (FRM) stars \citep{decressin2007} may be able to produce helium-rich ejecta. However the amount of helium ejected by massive AGB stars does not seem to be sufficient to raise Y to values $\gtrsim$0.35, and this could rule out them as progenitors of the extreme helium rich population. While the most massive FRM stars may provide the required quantities of helium but it is difficult to understand how they can be produced without any --- or, for $\omega$~Cen, without a considerable --- associated metal enhancement.

An appealing alternative may be represented by the SAGB stars ending their evolution as NeO WDs and, in this context, we have examined their possible role in the self-enrichment process, showing that they may provide the required high helium (see PDV08 for details). However, as can be seen from Fig. 1 of PDV08, the ejecta of the most massive SAGBs show a global CNO enrichment with respect to the initial value by a factor of $\simeq$4, due to the dredge-out process occurring at the second dredge-up phase\footnote{As in the intermediate-mass stars, the energy released by the core contraction after the central helium exhaustion induces the occurrence of the second dredge-up phenomenon, which reduces the mass of the H-exhausted core and increases the helium abundance in the envelope \citep[e.g.][]{vienna06,sp06}. However in some SAGB stars the second dredge-up is replaced by the so-called dredge-out phenomenon, in which the outer edge of a convective shell driven by the He-burning shell grows in mass and merges with the envelope \citep[see][for details]{ritosa99,sp06}.}. As a consequence, further observations of the very helium rich stars in GCs will allow us to better check this alternative scenario. In fact, if these quoted GCs show no evidence for this CNO enrichment, we may be able to conclude that at least the most massive SAGBs do not take part in the process of forming the second stellar generation in GCs and must evolve into ecSNe. In turn this hypothesis may help to explain also the high number ($\sim$ 500-1000) of neutron stars present in GCs, considering that the neutron star formed after the ecSN event would remain into the clusters, thanks to the small natal kick associated with such event \citep{ivanova2008}.

\section{The s-nucleosynthesis process in massive AGB and SAGB stars}

The so-called s-nuclei are formed via neutron exposures on iron-peak nuclei and, in terms of stellar sites, current views on the subject suggest the existence of two environments in which these nuclei can be synthetised \citep[e.g.][and references therein]{costa06}: one site is associated with low mass stars ($\sim$ 1.5-3$M_\odot$) during their AGB phase where the main neutron source is the $^{13}$C($\alpha$,n)$^{16}$O reaction (so-called main component of s-process); the other one is found in massive stars during their core He-burning phase and the main neutron source is the $^{22}$Ne($\alpha$,n)$^{25}$Mg reaction (so-called weak component of s-process).

However other kinds of stars, as massive AGB and SAGB stars ending their life as NeO WDs, could also contribute to the nucleosynthesis of s-species. In fact in these stars the physical conditions would be suitable to develop events of s-nucleosynthesis with features intermediate between the main component and the weak one. Specifically, s-nucleosynthesis episodes would take place during the AGB phase as in low-mass stars, but the main neutron source would be the $^{22}$Ne($\alpha$,n)$^{25}$Mg reaction as in massive stars \citep[e.g.][]{ritosa99, abia01}. Unfortunately little is known about the s-process in these stars \citep[][and references therein]{s07,firenze07} and a comprehensive study on this subject is needed not only in the framework of recent nuclear data and stellar models, but also in the light of the observed Rb-rich AGB stars \citep{science06}, whose Rb overabundances are not predicted by current s-nucleosynthesis theoretical models, which usually do not consider massive AGB and SAGB stars. 
 
In order to have quantitative information about the role of massive AGB and SAGB stars in the s-process, we are fully investigating the s-nucleosynthesis efficency in these stars, through a specifically written s-nucleosynthesis code (version modified for AGB and SAGB stars of the nucleosynthesis code described in \citet{costa06}; details will be provided in a forthcoming paper) coupled with the stellar evolution code ATON \citep{VD05}.

Preliminary results, referring to a Z=0.02 M=6$M_{\odot}$ massive AGB model, seem to indicate that the production of Rb is advantaged compared to other nearby s-elements (overproduction factor of Rb is $\sim 10-20$ times greater than the one of Zr, Y and Sr in the stellar layers where the neutron mass fraction raises to values $\sim 2\cdot 10^{17}$). On condition that the episodes of third dredge-up are efficient enough to reach such layers, the s-nucleosynthesis events in massive AGB and SAGB stars would be accountable for the Rb-rich AGB stars. 

A deeper analysis, involving also other masses, will allow us to better analyse and eventually confirm our preliminary interpretations.

\begin{acknowledgements}
M.L.P. and R.A.Z. thank the {\it Centro Siciliano di Fisica Nucleare e di Struttura della Materia} (CSFNSM) for financial support.
\end{acknowledgements}

\bibliographystyle{aa}

\end{document}